\documentclass{an}
\usepackage{graphicx}
\usepackage{url}
\usepackage{times}
\def\num{\hbox{N$^{\underline{o}}$}}
\Volume{00}              
\Year{2003}              
\Month{00}               
\Pagespan{000}{000}      

\begin{document}

\lhead[\thepage]{A.N. Koch et al.: Calibrating WFI photometry}
\rhead[Astron. Nachr./AN~{\bf} (2003) ]{\thepage}
\headnote{Astron. Nachr./AN {\bf } (2003) }

\title{A calibration map for Wide Field Imager photometry 
	\thanks{Based on observations carried out at the European
         Southern Observatory, La Silla, Chile; Program 60.A-9123A (MPIA guaranteed time)}
      }

\author{A. Koch \inst{1,2} \and M. Odenkirchen \inst{1} \and E.K. Grebel \inst{1,2}  \and  J.A.R. Caldwell \inst{1,3}}

\institute{Max-Planck-Institut f\"ur Astronomie, K\"onigstuhl 17, D-69117 Heidelberg, Germany 
\and Astronomisches Institut der Universit\"at Basel, Venusstr. 7, CH-4102 Binningen, Switzerland
\and Space Telescope Science Institute, 3700 San Martin Drive, Baltimore, MD 21218, USA }

\date{Received 2003 August 01; accepted 2003 October 10}

\abstract{
We present a prescription to correct large-scale intensity variations
affecting imaging data taken with the Wide Field Imager (WFI) at the 
MPG/ESO 2.2\,m telescope at the European Southern Observatory at La Silla
in Chile.  Such smoothly varying, large-scale gradients are primarily 
caused by non-uniform illumination due to stray light, which cannot be
removed using standard flatfield procedures. 
By comparing our observations to the well-calibrated,
homogeneous multi-colour photometry from the Sloan Digital Sky Survey 
we characterise the intensity gradients across the camera by
second-order polynomials.  The application of these polynomials to our data 
removes the gradients and reduces the overall scatter.  We also
demonstrate that applying our correction to an independent WFI dataset 
significantly reduces its large-scale variations, indicating that our
prescription provides a generally valid and 
simple tool for calibrating WFI photometry.
}
\correspondence{A. Koch ({\tt koch@astro.unibas.ch})}

\maketitle

\section{Introduction}
The quality of any astronomical imaging system is limited by several factors. For instance, the detectors, which nowadays are predominantly CCDs, 
show (small-scale) variations in quantum efficiency. Such shortcomings are intrinsic to the CCD camera and can usually easily be corrected during the reduction process, 
especially if they do not vary with time. This is generally achieved by, e.g., applying more or less sophisticated flatfield calibrations. 
Certain large-scale effects may, however, vary with time and telescope position and are much more difficult to handle. In particular, in complex instruments with many optical
surfaces, stray light can hardly be avoided.  This implies that neither science frames nor flatfield exposures are illuminated uniformly. 

The Wide Field Imager camera (WFI) at the MPG/ESO 2.2\,m telescope at the
European Southern Observatory (ESO), La Silla, Chile, is the only
southern-hemisphere
wide-field, high-resolution, CCD imaging facility available to the
general ESO community.  The WFI is being used for large public surveys
such as the ESO Imaging Survey (EIS; see 
\url{http://www.hq.eso.org/science/eis/}),
for other small and large programs preparing observations with the Very
Large Telescope (VLT) at ESO's Cerro Paranal Observatory, and for numerous 
independent science programs.
The WFI has been available to the general ESO community since 1999 
(Baade et al. 1999) and has repeatedly been reported to show 
significant 
large-scale spatial gradients in photometry across the entire field of view (see, e.g., Manfroid et al. 2001a), and  across each of its eight chips individually. 
In order to obtain homogeneous, reliable, and reproduceable photometric results, these effects need to be carefully and thoroughly corrected.
Therefore, it would be desirable to devise a general method that can be applied by WFI users to their photometry, both for currently ongoing projects
and for a proper exploitation of the large amount of WFI data already in
the archives. 

The evaluation of presence and magnitude of large-scale photometric variations necessitates a comparison of observational data against a calibrated sample. 
Generally, fields containing a relatively small number of standard stars are used for this purpose, e.g., Landolt fields (Landolt 1992). 
However, in order to characterise spatial effects across the entire field of view and to achieve a well-sampled, reliable calibration of wide-field photometry, ideally one 
needs to observe the same standard field on each of the single CCDs, and to repeat this
procedure for a number of fields at different airmasses. In practice, this procedure is immensely time-consuming.
It is a lot more efficient to calibrate the observed data against well-defined datasets covering the same or larger areas down to similar magnitude limits. While such data are
not normally readily available, our work benefited from the fact that all areas observed by us coincide with fields already surveyed by the Sloan Digital 
Sky Survey (SDSS) that are now publicly available as part of the SDSS Early
Data Release (Stoughton et al. 2002) and SDSS Data Release 1 (Abazajian et al. 2003).   

The SDSS is a large CCD sky survey with the ultimate goal to provide accurate deep multi-colour imaging and spectroscopy 
for up to one quarter of the celestial sphere. Since SDSS imaging is being carried out in driftscan mode (York et al. 2000), the resulting photometry is highly 
homogeneous and uniform.
The availability of the SDSS data allowed us for the first time to compare WFI photometry directly against a dense grid of local ``standards'' and thus to directly 
calibrate the science frames.

This paper is organised as follows: Section 2 summarises the observations and basic reductions. In Section 3 we give an overview of the method used for the photometric
calibration. The results are presented in Section 4. In Section 5 we apply
our prescription to an independently obtained WFI data set. 
The discussion and summary of our results are presented in Section 6.  
%

\section{Observations and reduction}
During two nights in May 2001 we used the WFI to perform observations of the faint and sparse globular cluster Palomar 5 
($\alpha (J2000) =  229 \fdg 019,\, \delta (J2000) = -0\fdg 121$, see, e.g., Odenkirchen et al. 2001, 2003). 
In this run we observed three fields.  One field targeted the cluster's center (labelled F1). The second field was centered on a density node 
located in one of the tidal tails (F2).  The third field was chosen as a comparison field (F3) at 1$\fdg$5 distance from the cluster center, away from the 
cluster's tidal tails.
The location of these fields is depicted in Figure 1.
These data were obtained in order to study the luminosity function of the globular cluster and its tidal tails, which will be described in a subsequent paper (Koch et al., in prep.; see also Koch 2003). In the
current work we explore instrumental effects affecting WFI photometry. For this purpose the choice of the fields is not important as long as excessive crowding is avoided and
overlap with the SDSS is guaranteed.
\begin{figure}[t]
\resizebox{\hsize}{!}{\includegraphics{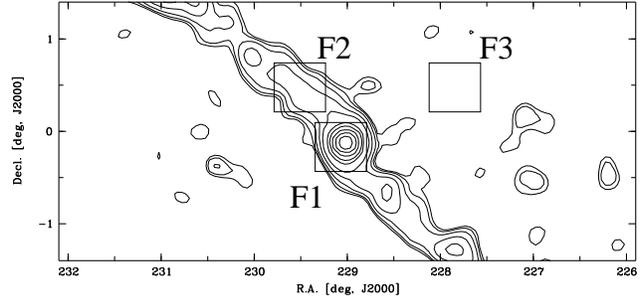}}
\caption{The observed fields F1, F2, and F3 -- marked by squares -- overlaid on a contour plot of Palomar 5's density distribution. The central density concentration at $\alpha =
229\degr,\,\delta=0\degr$ (J2000) marks the cluster center, from which two tidal tails are emerging (Odenkirchen et al. 2001).}
\label{f:fields}
\end{figure}

The observations were taken both in the V and R filter, where each of the fields was exposed five times. An observing log is presented in Table 1. 
\begin{table}[b]
\begin{center}
\caption{Observation log}
\begin{tabular}{c c c c c}
\hline
\hline
Date & Field & Filter & Exposure & $\num$ of \\
     &	     &        & time [s] & exposures \\
\hline
2001 May 17 & F2 & V  & 900 & 5\\
 '' & F2 & R & 600 & 5\\
 '' & F3 & V & 900 & 5\\
2001 May 18 & F3 & R & 600 & 5\\
 ''& F1 & V & 900 & 5\\
 '' & F1 & R & 600 & 5\\
\hline
\end{tabular}
\label{Obs}
\end{center}
\end{table}
The exposures were offset against each other by approximately 15$\arcsec$ in the vertical and horizontal direction in order to cover 
the gaps between the single WFI CCDs. In addition, bias and twilight flatfield exposures were obtained in each night. 
The seeing in the first night was approximately 1$\farcs$1 and improved to 0$\farcs$75 in the second night, whereas the airmass was 1.2 on average.
The relative constancy of the airmass as well as the small and
roughly linear colour dependence of atmospheric extinction in the V and R 
filters (Roberts \& Grebel 1995) alleviates the need for additional
atmospheric extinction corrections in the subsequent reduction. 
The observations were carried out during new moon, hence moon light did not provide an additional, time varying stray light component.

The obtained raw data files (both science, bias and flatfields) were split into eight single images, each corresponding to one individual CCD chip. Thus, during all of the subsequent reduction steps, each of the
chips was treated as a separate frame. The standard reduction has been carried out using the IRAF package. As a first step
readout bias was removed to first order by subtracting a fit of the overscan region from the frames. 
Any residual bias was then subtracted using the average resulting from our 30 bias frames.
Finally, flatfield calibrations was carried out using the qualitatively best of the observed twilight flats, where an individual 
scale factor was determined for each single chip by taking the mean flatfield value over the whole CCD chip, clipping the frame at four standard deviations of this value and 
iterating until the mean got stable. 
Hence, having normalised the flatfield on each CCD to 1 by using the respective mean determined above, the gain differences between single chips 
were preserved.
Neither dark current nor fringing causes any considerable effect in WFI observations made in V and R so that these effects have not been corrected for. 

Afterwards, we performed aperture photometry on the reduced frames using the DoPHOT program (Schechter et al. 1993) for object finding, classification, and photometry.
The digital aperture for this procedure was chosen as 25$\times$25 pixels or 6$\arcsec$ square.  
%

\section{Setting up a photometric comparison grid}

In order to quantify the variations in WFI photometry across the camera, we compare our WFI data to the photometrically and spatially 
homogeneous SDSS system. 

\subsection{The Wide Field Imager}
The WFI is a mosaic of eight rectangular CCD chips arranged in 2 rows \`a 4 CCDs, where each chip consists of 2046$\times$4098 pixels at a pixel scale of 0.238
arcsec pixel$^{-1}$ (Baade et al. 1998). Additionally, 
there is a tracker CCD, which is located east of the science mosaic and is used for guiding of the telescope.  
It is of identical design to the imaging CCDs. The whole camera provides a field of 
view of 34$\arcmin\,\times\,$34$\arcmin$.  Due to the physical separation 
between the chips,  
the actual area coverage is 95.9\%. The sizes of the gaps between the chips are 23$\arcsec$ in vertical direction and 7$\arcsec$
in horizontal direction. 
Figure 2 shows an overview of the arrangement of the mosaic.

\begin{figure}[h]
\resizebox{\hsize}{!}{\includegraphics[clip]{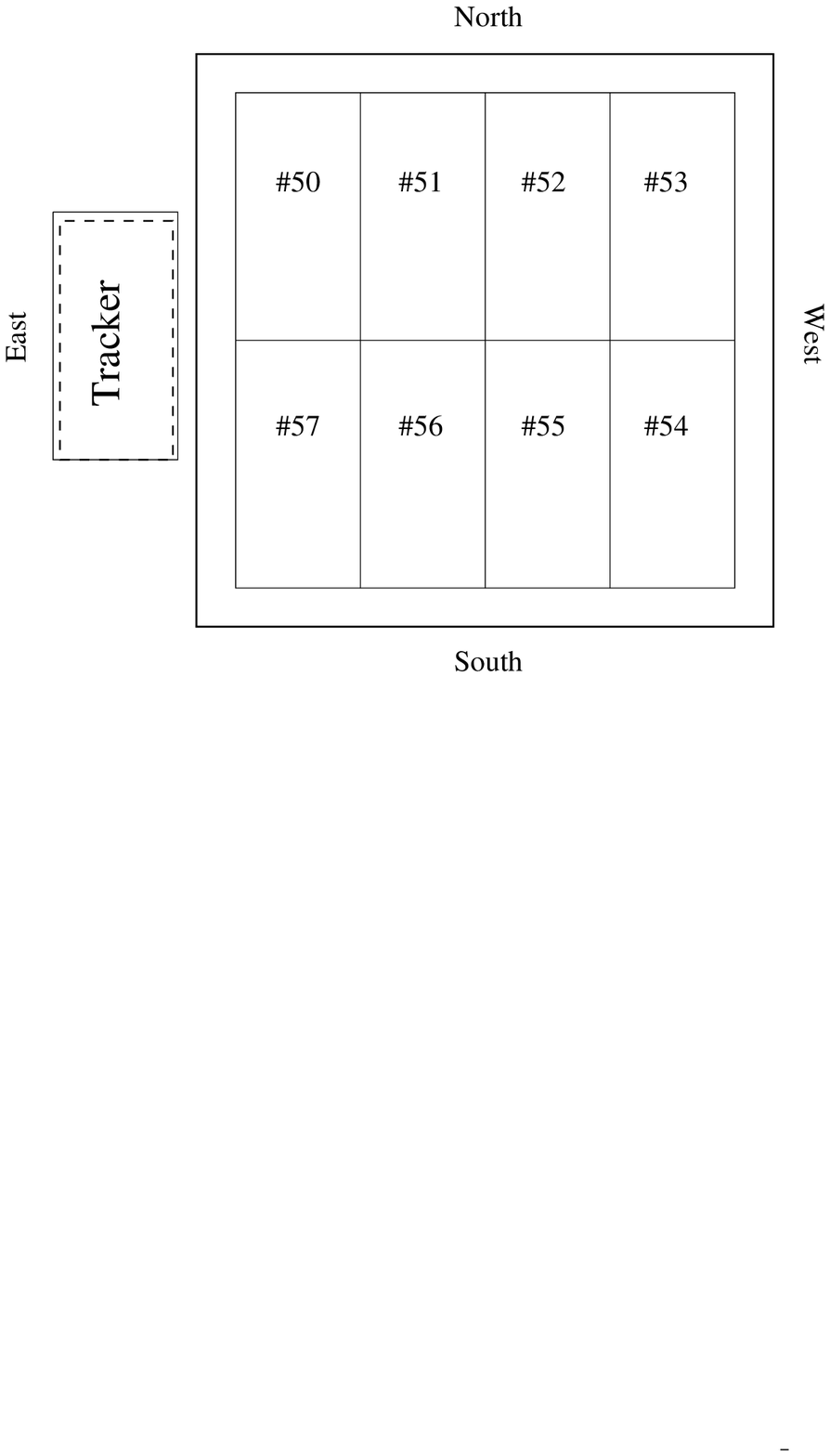}}
\caption{The arrangement of the eight WFI CCDs, shown as a projection of the detector plane on the filter plane (outer square). The numbers preceded by a
hash follow the official ESO nomenclature. (Adapted from the WFI User Manual.)}
\label{WFI}
\end{figure}
%
%

\subsection{The Sloan Digital Sky Survey}

For an overview of the SDSS, we refer to the on-line SDSS Project 
Book\footnote{(\url{http://www.astro.princeton.edu/PBOOK/welcome.htm})} 
and to reviews such as Grebel (2001).  
The data for this multi-colour imaging and spectroscopic survey are obtained at the 2.5\,m telescope of the Apache Point Observatory (New Mexico, USA). 
The camera used for the imaging part of the SDSS consists of a mosaic of 5$\times$6 CCDs, operated in driftscan mode (Gunn et al. 1998; York et al. 2000). 
Two contiguous scans cover a field of view of 2$\fdg$5 on the sky.
The SDSS filter system is a modified Thuan-Gunn filter set 
labelled u, g, r, i, and z, which is described in
more detail in Fukugita et al. (1996), Hogg et al. (2001), 
and Smith et al. (2002). 
SDSS magnitudes used in our analysis are based on the PSF photometry from the photometric pipeline as described in 
Lupton et al. (2001).
Figure 3 shows a plot of the five SDSS transmission curves compared to the standard broad-band filter system used with the WFI. These U, B, V, R$_{\mathrm{C}}$ 
and I$_{\mathrm{C}}$ filters are similar, but not fully identical to the Johnson-Cousins UBVR$_{\rm C}$I$_{\rm C}$ system (Johnson \& Morgan 1953; Cousins 
1978; see Girardi et al. 2002 for a discussion of the differences).   

\begin{figure}[h]
\resizebox{\hsize}{!}{\includegraphics{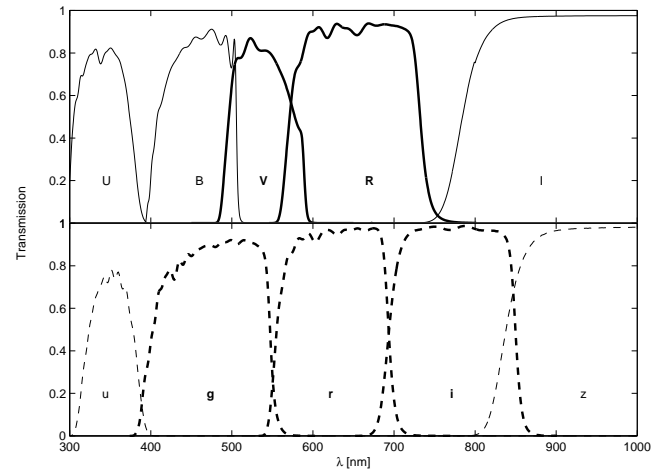}}
\caption{The transmission curves of the Johnson-Cousins
filters used for the WFI {\em (top panel)} and in the SDSS {\em (bottom panel)}. Thick lines indicate the
filters that are used in our transformations.}
\label{filters}
\end{figure}

\subsection{Transformation between WFI and SDSS system}

Objects within the relevant region of the equatorial stripe ($\delta\,(J2000)\,\sim\,0\degr$), comprising our three observed fields,  were extracted from the SDSS database
by means of the SkyServer tool (Szalay et al. 2001). These are part of the SDSS Early Data Release (EDR, see Stoughton et al. 2002). 
Only objects classified as stars by the SDSS were considered in our sample. Positions on the WFI frames were astrometrically transformed into the equatorial system
($\alpha,\,\delta$) using 
iteratively the IRAF routines {\em geomap} and {\em geotran} based on a sample of approximately 30 uniformly distributed 
reference stars per exosure for each CCD. Afterwards each single exposure
was matched by position against the SDSS sample. 
This yielded a total number of about 500
stars on each CCD of each exposure in each of the three observed fields.
In order to avoid saturated objects (in the SDSS) and stars with larger photometric errors (both for the WFI and SDSS photometry) we selected only stars
satisfying  $16\fm3\le$\,r\,$\le 21\fm7$ in order to determine the transformation relations between the two systems.
After all this, approximately 200 stars remained per chip and per exposure 
of each respective field to be used for the photometric transformations.

Since there is an overlap of the V  and R transmission curves with the SDSS filters g, r, i, it appears useful to define
the transformation equations between WFI instrumental magnitudes\footnote{Henceforth
WFI {\em instrumental} magnitudes will be denoted by an asterisk.} and the SDSS magnitudes as
\begin{eqnarray}
R^{\ast}&\,=\,&r+\alpha_R\,(g-r)\,+\,\beta\,(r-i)\,+\,c_R, \\
V^{\ast}&\,=\,&g+\alpha_V\,(g-r)\,+\,c_V,
\end{eqnarray}
with colour coefficients $\alpha,\,\beta$ and zeropoints $c$. It
was found that quadratic terms such as $(g-r)^2$ have negligible influence on the quality of the transformations so that these were not included.

The entire transformation according to Eqs. (1),(2) was carried out for each of the eight individual frames of each single exposure. 
The values for the coefficients show little scatter from exposure to exposure, amounting to approximately 1\%. 
Their error-weighted mean values are listed in Table 2, where the 
1$\sigma$ fitting error 
for the coefficients is of the order of 0.005.
\begin{table*}[t]
\begin{center}
\caption{Coefficients of the transformation equations (1) and (2), averaged
over all exposures.}
\begin{tabular}{l r r r r r r r r}
\hline
\hline
CCD chip & \#50 & \#51 & \#52 & \#53 & \#54 & \#55 & \#56 & \#57 \\
\hline
$\alpha_R$ & $-$0.017	& $-$0.006 &     0.026&    0.006  &   0.018  &    0.001 & $-$0.004 &  0.011 \\
$\beta$    & $-$0.169	& $-$0.221 & $-$0.254 & $-$0.208  & $-$0.235 & $-$0.223 & $-$0.214 & $-$0.249 \\
$c_R$      & $-$30.349	& $-$30.309& $-$30.179& $-$30.378 & $-$30.339& $-$30.353& $-$30.309& $-$30.219 \\
$\alpha_V$ & $-$0.489	& $-$0.491 & $-$0.499 & $-$0.487  & $-$0.486 & $-$0.504 & $-$0.491 & $-$0.493 \\
$c_V$      & $-$30.207	& $-$30.343& $-$30.090& $-$30.333 & $-$30.312 & $-$30.299& $-$30.229& $-$30.106 \\
\hline
\end{tabular}
\end{center}
\label{coeff1}
\end{table*}

%
\section{Residuals of the transformation}

\subsection{Measured residuals}
The residuals of the transformations are defined as follows:
\begin{eqnarray}
\varepsilon_R\,&=&\,R^{\ast}\,-\,(
r+\alpha_R\,(g-r)\,+\,\beta\,(r-i)\,+\,c_R )\\
\varepsilon_V\,&=&\,V^{\ast}\,-\,( g+\alpha_V\,(g-r)\,+\,c_V ).
\end{eqnarray}
These allow judging the quality of the best-fit solution and to determine systematic differences between WFI and SDSS photometry.
In order to visualise the spatial variations of the residuals, a plot of $\varepsilon$ versus
location of the objects on the CCDs is presented in Figure 4.

\begin{figure}[h]
\resizebox{\hsize}{!}{\includegraphics[clip]{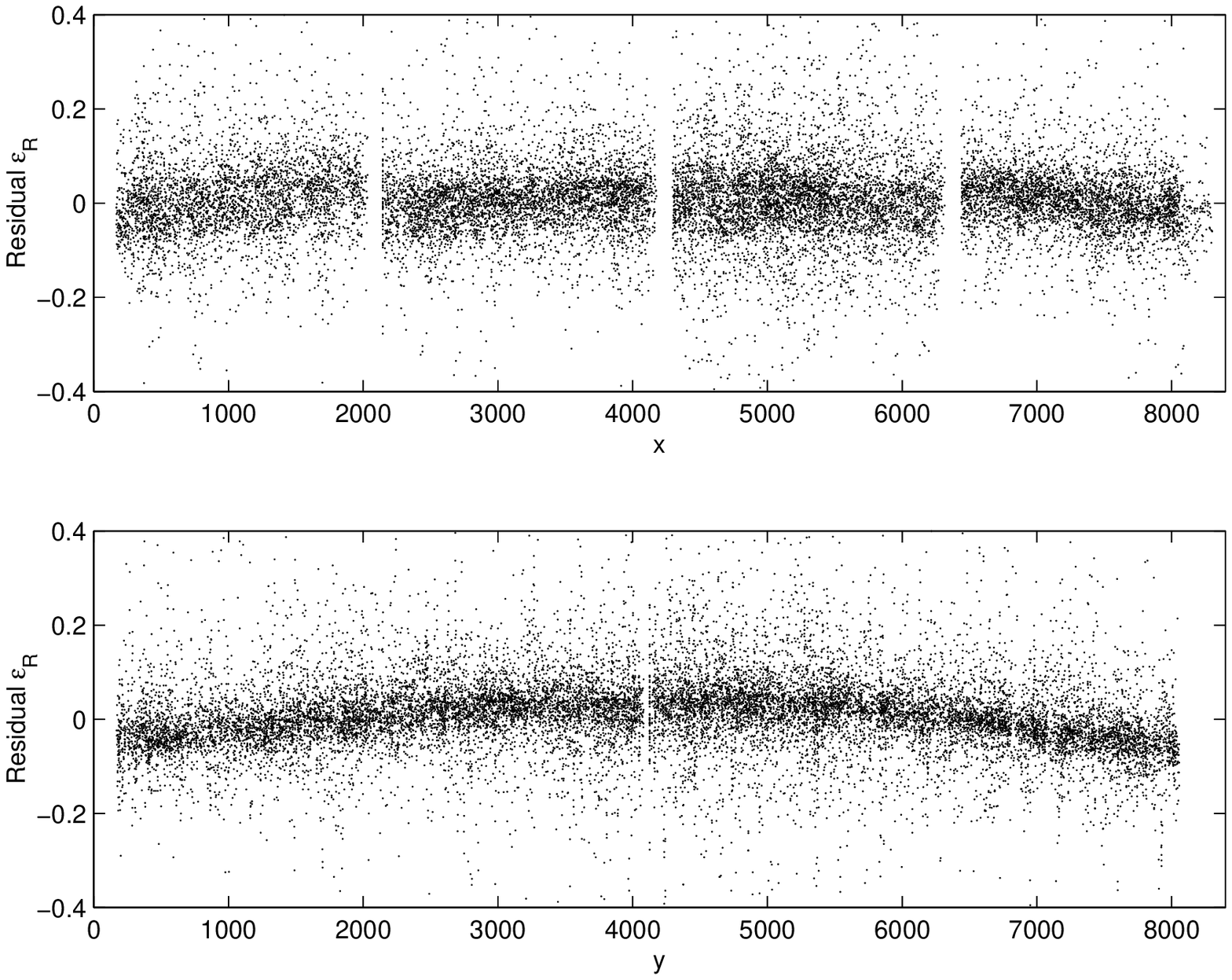}}
\caption{Residuals for the $R$-band magnitude of 17754 objects as defined in equation (3) versus location on
the WFI. Gaps and gradients are clearly visible.
The V-band residuals exhibit a similar spatial pattern to the R trends shown here. 
Zeropoint differences reach peak-to-valley amplitudes of 0.19\,mag both in R (cf. lower panel) and V, whereas the r.m.s. scatter is about 0.08\,mag for both filters.
}
\label{measured}
\end{figure}
The coordinate system is oriented such that the origin is located at the bottom left
corner of the camera with the axes increasing (in units of pixels) from bottom to top (y) and left to right (x), respectively.

Two outstanding features are distinctly  to be seen: First, the distribution of the residuals is not flat within each chip.
Instead, strong gradients occur, which appear parabolic in shape.  Consequently, objects
measured near the center of the camera appear fainter than in its outskirts. The overall observational scatter (including random and systematic errors) of the 
residuals amounts to approximately 0.08\,mag on each single CCD chip. 

While the high-frequency pixel-to-pixel variations that are generally handled by usual flatfields are caused by variations in quantum efficiency, 
large-scale gradients can originate because of 
non-uniform illumination, e.g., due to reflections inherent to the WFI camera (i.e., scattered light on the instrument).
Such a redistribution of light by the optical system affects the flatfield as well, corrupting it significantly. Mainly, this non-uniform illumination causes more light  to
reach the center of the detector, an effect that is known as sky concentration (Andersen et al. 1995).
Thus, when applying the multiplicative flatfield correction, the flux of stars near the center of the mosaic will be artificially reduced and one obtains an image 
that is falsified on a large scale (Selman 2003, in prep.\footnote{\url{http://www.ls.eso.org/lasilla/sciops/2p2/E2p2M/WFI/zeropoints/zpmap/index.html}}).

Secondly, there are offsets between the single chips. But considering our reduction procedure, which treated each chip individually, this is not surprising. In particular,
during the flatfielding each CCD chip was normalized to a mean intensity level determined for this specific chip, resulting in intensity level offsets from chip to chip.

\subsection{A model for the spatial systematics}

The shape of the curves in Figure 4 implies a low-order polynomial variation of the residuals with position.
Thus we modelled the residuals by a complete polynomial of second order, i.e., 
\begin{eqnarray}
\varepsilon (x,y)\,=\,A\,x^2\,+\,B\,y^2\,+\,C\,xy\,+\,D\,x\,+\,E\,y\,+F 
\end{eqnarray}
to describe the photometric gradients, 
where A to F denote the fit parameters. These parameters were determined by fitting the model (5) to the observed residuals (Eqs. 3 and 4) by means of a non-linear least
squares algorithm.  The residuals were weighted with 
\begin{center}
\begin{equation}
w\propto\frac{1}{N\,\sigma^2} ,
\end{equation}
\end{center}
where $\sigma$ is the photometric uncertainty and N the number density of objects
around each respective
star.  
The latter was introduced in order to avoid that the fit of the model be dominated by stars in the region of the cluster and thus
spatially biased.
The output data from all exposures on the same chip were now combined to form a sample of numerous photometric data points with excellent spatial sampling.  
The outer 40$\arcsec$ of the CCD chips were excluded from the fit, since in these regions one expects strong variations (Manfroid et al.
2001b). This leaves a total number of approximately 2200 stars per CCD chip, or 17754 stars for the whole camera, to be used in the fit. Since every
physical star can appear up to five times at dithered locations, the sample increased nearly fivefold compared to single exposures.
The resulting coefficients of the best-fit model are listed in Table 3\footnote{This table is available in electronic form from \url{http://www.mpia-hd.mpg/SDSS/data}}
. 

\begin{table*}[t]
\begin{center}
\caption{Coefficients of the polynomial equation (5). The {\em top} part lists values for the R filter, the {\em bottom} part shows the coefficients from the 
V transformation.}
\begin{tabular}{c r r r r r r r r }
\hline
\hline
Coefficient R & \#50 & \#51 & \#52 & \#53 & \#54 & \#55 & \#56 & \#57 \\
\hline
A ($\times\,10^9$) & $-$16.1 &    0.43  & $-$1.03 & $-$1.47   &  4.66 & $-$15.4 & $-$9.48 & $-$5.15 \\
B ($\times\,10^9$) & $-$9.30 & $-$8.00  & $-$6.27 & $-$7.84 & $-$1.94 & $-$7.12 & $-$9.20 & $-$5.25 \\
C ($\times\,10^9$) & $-$0.28 &    1.49  &    2.57 &    2.14 &    6.63 &    4.01 &    1.26 & $-$0.69 \\
D ($\times\,10^5$) &    6.11 &    0.56  & $-$0.64 & $-$2.57 & $-$5.27 &    1.20 &    2.95 &    4.29 \\
E ($\times\,10^5$) &    0.32 &    0.18  & $-$0.28 &    0.86 &    1.94 &    4.62 &    5.63 &    4.32 \\ 
F ($\times\,10^2$) &    0.47 &    3.11  &    4.37 &    4.71 &    0.23 & $-$5.50 & $-$8.69 & $-$9.55 \\  
\hline
Coefficient V & \#50 & \#51 & \#52 & \#53 & \#54 & \#55 & \#56 & \#57 \\
\hline
A ($\times\,10^9$) &    20.9 & $-$12.4 & $-$6.54 &    3.75 &    9.05 & $-$9.13 & $-$8.59 &    29.3 \\
B ($\times\,10^9$) & $-$5.00 & $-$14.0 & $-$8.00 & $-$6.44 & $-$3.79 & $-$8.29 & $-$10.8 & $-$3.46 \\
C ($\times\,10^9$) & $-$1.38 & $-$3.66 &    3.07 &    0.87 &    0.71 &    3.75 &    2.33 &    0.53 \\
D ($\times\,10^5$) & $-$2.00 &    4.43 &    0.19 & $-$4.31 & $-$4.80 & $-$0.04 &    2.77 & $-$3.40 \\
E ($\times\,10^5$) & $-$0.57 &    1.97 &    0.19 &    0.89 &    2.76 &    5.15 &    6.79 &    3.22 \\ 
F ($\times\,10^2$) &    3.28 &    1.14 &    4.06 &    4.75 & $-$0.30 &  $-$5.61& $-$10.6 & $-$5.71 \\  
\hline
\end{tabular}
\label{coeff2}
\end{center}
\end{table*}

In order to get an estimate of the pure global properties of the overall gradients we removed the intensity level offsets caused by
the individual flatfielding of each CCD chip. Assuming that illumination effects such as stray light are the reason for the structures we found and that the CCDs themselves are basically homogeneous, 
removal of the intensity offsets should leave the mere global illumination pattern. Thus, to bring all chips (with hindsight) to a common level, a zeropoint adjustment was 
calculated as follows.
Chip \#53 was chosen as reference, since it appears to be the 
cosmetically best chip. 
The mean values of the residuals (Eqn. 5) at the edges of adjacent CCDs were adjusted successively.
To be consistent, this was done iteratively at the horizontal and the vertical edges.
It turns out that the values for these additive offsets are of comparable size to the fit 
constants F in \mbox{Eqn. 5.} 

The peak-to-valley amplitudes of the photometric variations given by the model range from 0.11\,mag (\#52) to 0.19\,mag (\#50) for the R filter corrections 
and likewise from 0.11\,mag (\#54) to 0.19\,mag (\#51) on the V map. 
A plot of the resulting, offset-modified calibration map according to Eqn. 5 is shown in Figure 5.
An obvious benefit of our procedure is (per constructionem) that the residual steps at the CCD edges vanish, resulting in smoothly varying overall gradient maps.
The appearance of such a smooth structure shows that in fact stray light on a large scale dominates the systematic variations in photometry.

There is a close overall similarity between the maps of the R  and V filters.
The residuals on the V map appear more centrally concentrated than those
on the R map.  An additional difference is the occurrence of a prominent feature in the V map, which is distinctly visible on chips \#50 and \#57: There is a bumpy structure or band of higher 
brightness along the vertical axis.
This is a well-known problem and stems from stray light from the tracker CCD. 
Despite the use of shielding baffles, light from bright stars on the tracker CCD may be reflected onto the eight main CCDs of the WFI camera.  
The tracker CCD is used with auxiliary filters
(dashed rectangle in Fig.\ 2),
which have a central wavelength comparable to that of the science filters currently used, but a much wider bandpass 
than the broad-band science filters.   Hence those science filters whose
passbands overlap with the auxiliary filter will be affected by this
additional source of stray light.  In our case, the V-band map exhibits the
signature of this effect, while the R-band map does not\footnote{The central
wavelengths of the V and R filters are 539.6 nm (FWHM(V) = 89.4 nm) and 
651.7 nm
(FWHM(R)=162.2 nm), respectively.}.  Furthermore, this excess of light
on the science mosaic depends on the position and apparent luminosity of
bright sources on the tracker CCD, making it difficult to correct 
for these effects by simply dividing by the flatfield exposures. 

\begin{figure*}[t]
\begin{center}
\resizebox{12cm}{!}{\includegraphics[clip]{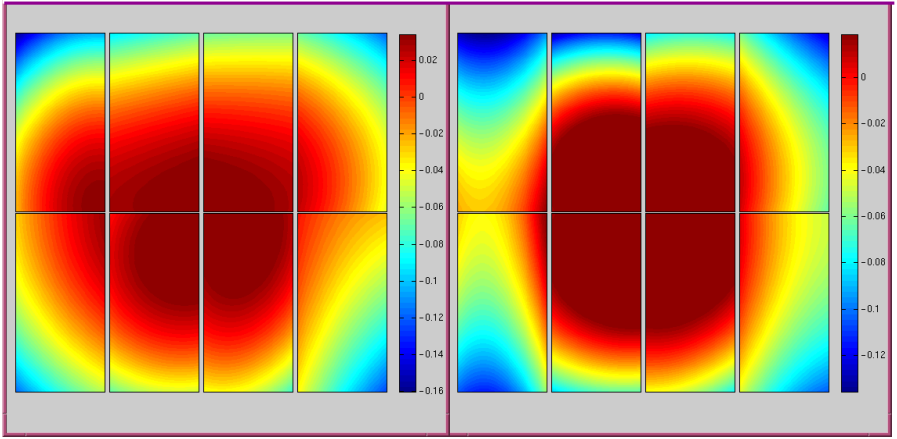}}
\caption{Our best-fit second-order calibration map after application of the mean offsets that were derived to correct for different flatfield scale factors.
Note that all chips were processed individually. The R-band map is shown in 
the left panel, while the right panel displays the V-band map.
The colour scale is in magnitudes.  R-band exposures are known to be affected by 
vignetting in the outermost corners of the camera.}
\end{center}
\end{figure*}

\subsection{Remaining residuals}

The benefit of the model becomes clear if one compares pre- and post-fit residuals. Figure 6 shows the post-fit residuals 
\mbox{$\varepsilon_{Obs}\,-\,\varepsilon_{Model}$} plotted versus position as in 
Figure 4. 
(Here $\varepsilon_{Obs}$ are the residuals as measured from Eqs. 3 and 4, and $\varepsilon_{Model}$ are those derived from Eqn. 5.) First, after applying Eqn.\ 5 to our data the offsets at the chip separations have essentially vanished.
The second and more important outcome 
is a considerable flattening of the large-scale structure in the residuals, which suggests that our model is successful at significantly reducing
gradients. Nevertheless, there are still a few slight variations left, especially at the very
edges, where the model was extrapolated instead of derived from the fit.  

\begin{figure}[h]
\resizebox{\hsize}{!}{\includegraphics[clip]{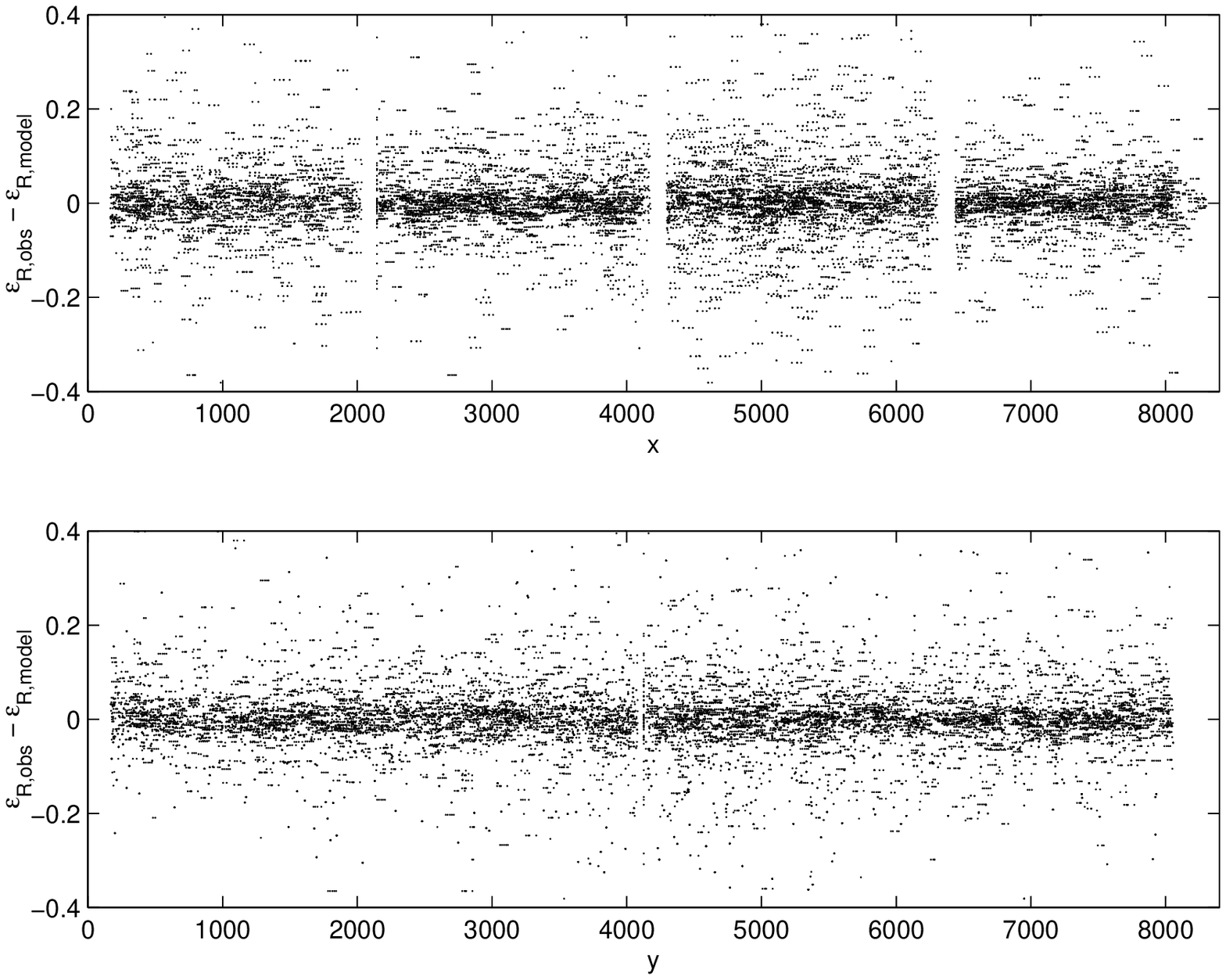}}
\caption{Statistical residuals (R filter) remaining after the model (Eqn. 5) was
subtracted from the residuals in Figure 4. Points appearing fivefold and shifted horizontally are the same physical objects in the five dithered observations of each field.
The overall r.m.s. scatter was reduced to 0.06\,mag in R and V.
}
\label{remain}
\end{figure}

\section{Independent tests}

Our calibration map provides valid corrections for the large-scale variations in the set of observations of Palomar 5 (and the other
fields) for which it has been constructed. In order to test whether it is also useful to improve other observations, the coefficients were applied to an entirely independent
photometric dataset. This dataset consisted of WFI observations of the globular cluster NGC 6934, which we obtained in V and R on 2000 September 5.

In contrast to the data on Palomar 5, the flatfielding here was carried out for the entire mosaic simultaneously.
That is, the mean value of the entire flatfield exposure was determined, thus normalising the whole mosaic to a common gain.
Zeropoint offsets between the CCDs are now removed when dividing the science frames by the flatfield mosaic, 
 bringing all eight chips to the same intensity levels to begin with.
The top panel of Figure 7 shows differential photometry 
$\Delta R$ obtained by artificially shifting two exposures against each other by one chip size in y-direction ($\approx\,16\arcmin$).
That means, $\Delta R$ equals the difference between magnitudes of stars measured on the lower CCD panel (chips \#54 to \#57) and those of the same stars observed on the 
upper panel (\#50 to \#53).

Due to the simultaneous reduction of all CCDs in the mosaic, offsets 
arising from differing flatfield scale factors as in the case of Palomar 5
are not obvious.  
But again, there are spatially dependent variations in the photometry to be seen, similar to the gradients in the fields around Palomar 5.  
To correct for these large-scale gradients, we applied the polynomials with
the coefficients given in Table 3, omitting the additive scale factor F. 

\begin{figure}[h]
\framebox[\hsize]{\parbox{\hsize}{
Fig. 7 is available in jpg format from 

 http://lanl.arxiv.org/ps/astro-ph/0310301}}
\caption{Residuals in differential photometry (R filter) of NGC\,6934. {\em Top panel}: Differences as a function of position on the WFI CCDs.
{\em Bottom panel}: Residuals
remaining after having applied our correction model. The overdensity on chips \#51/\#56 is due to the globular cluster. 
In this independent test the r.m.s. scatter reduced from  0.07\,mag (cf. top panel) to about 0.04\,mag (bottom panel).
}
\end{figure}

After subtraction of the model both gradients and overall scatter turned out to be visibly reduced, as is shown in the bottom panel of Figure 7. 
The small residual variations at the outer rims of the CCDs seem unavoidable when using a low-order model, which only accounts for effects in the inner regions of the individual CCDs. 
Considering that the observations of NGC 6934 and their reduction were performed entirely separately from the Palomar 5 dataset, it is encouraging that our calibration method yields such a
good result in terms of strongly reduced systematic variations.  This indicates that
it can be generally applied to other datasets to correct large-scale gradients.

\section{Discussion and Summary}

As many wide-field imaging systems, CCD photometry obtained with the
WFI suffers from large-scale intensity gradients caused by inhomogeneous
illumination and stray light.  These gradients cannot be removed by 
standard techniques such as the application of dome flats or twilight
flats.  Previous methods 
that were proposed
to calibrate WFI photometry include
superflats (see, e.g., Clowe \& Schneider 2001 or Alcal\'a et al. 2002),
the creation of calibration maps shifting exposures with respect to each
other (Manfroid et al. 2001a, Selman 2003), and the determination of 
corrections through observations of Landolt (e.g., Landolt 1992) and Stetson 
standards 
 (Stetson 2002)
in each of the CCD chips.

Superflats have the advantage of providing corrections for large-scale
illumination effects determined from the science exposures themselves.
However, median sky flats, which are often used as superflats to ensure a flat sky background,  
encounter the same photometric problems as common sky flats.
Thus it is argued that application of such superflats can even deteriorate data by increasing photometric errors 
(Manfroid et al. 2001b).  
Another
drawback of this method is that it can only be applied in cases 
where a sufficiently high number of well-exposed science frames has
been obtained to ensure sufficient signal to noise. 
An additional disadvantage
is that it does not work well in cases where the majority of the science
exposures contain extended objects or where the observed fields are very
crowded.  Especially when observations are obtained in queue scheduling
mode, it may be difficult to obtain science frames suitable for the
creation of well-defined superflats.
 
Depending on the science program, obtaining images that are shifted by
half a WFI field with regard to the previous exposure may be practicable
(see also Section 5).  However, this will provide primarily only a relative
calibration of the two rows of chips against each other.

As discussed in Section 1, observing Landolt (and Stetson) standards in
each chip will ensure a good characterisation of the CCD sensitivity,
but is limited by two constraints:  Firstly, the number of standard stars
per chip remains comparatively small and may not provide as good a spatial
characterisation as desirable.  Secondly, depending on the science program
the amount of time needed for these calibration observations may be 
prohibitive. 

Therefore, we have attempted to devise a generally applicable method for 
the correction of large-scale spatial variations across the WFI.  This
method relies on the comparison with spatially very well-sampled, very
homogeneous multi-colour SDSS photometry, which we employ as tertiary
standards.  Our derived calibration maps are based on observations of 17754
stars.  We have demonstrated that second-order polynomials resulting from
a fit to the observed large-scale variations provide a very good  
correction by essentially removing the gradients and by significantly 
reducing the scatter in the residuals of our data.  We have also shown
that our correction relations are applicable to independently obtained
and separately reduced WFI data sets, significantly reducing gradients 
and scatter in the photometry.

Observers wishing to remove large-scale gradients from their WFI data
without having access to, e.g., standard star observations in each chip,
will be able to considerably reduce these gradients by subtracting
our position-dependent polynomials (Eqn. 5, Table 3) from their data. 
We expect that our prescription will also be valuable for the
exploitation of the wealth of archival WFI data.  
Our correction model was determined for observations in the widely used
V and R filters.  The overall similarity of the spatial variations in
these two filters and the nature of stray light effects may indicate 
that our corrections are also applicable to broad-band data obtained in
other filters, but this still needs to be verified.  Our correction
model yields global corrections, but we emphasise
that fine tuning for small-scale variations
and for possible differences in the illumination pattern (e.g., due to
bright moon light) will still be required.
Moreover, the stability of any calibration map can only considered as generally valid
unless there is no significant change in the optical setup of the telescope\footnote{
For instance, our coefficients cannot correct data taken in the 
period of September 2002 until December 2002, where an M1 baffle re-engeneering was performed at the 2.2\,m telescope, 
cf. \url{http://www.ls.eso.org/lasilla/sciops/2p2/E2p2M/WFI/ConfigurationControl/}.}.

The SDSS provides an excellent database for evaluating the photometric
quality and systematic effects in other data sets due to its homogeneity,
wavelength and area coverage.
We encourage WFI users (as well as users of other wide-field
cameras) to pursue similar calibrations by exploiting such 
publicly available multi-colour driftscan
surveys like the SDSS whenever possible.  Moreover, customised 
prescriptions similar to the one presented here will be useful for
the correction of large-scale illumination effects in wide-field
cameras used at other telescopes or observatories.


\begin{acknowledgements}
Funding for the creation and distribution of the SDSS Archive has 
been provided by the Alfred P. Sloan Foundation, the Participating 
Institutions, the National Aeronautics and Space Administration, 
the National Science Foundation, the U.S. Department of Energy, 
the Japanese Monbukagakusho, and the Max Planck Society. The SDSS 
Web site is \url{http://www.sdss.org/}. 

The SDSS is managed by the Astrophysical Research Consortium (ARC) 
for the Participating Institutions. The Participating Institutions 
are The University of Chicago, Fermilab, the Institute for Advanced 
Study, the Japan Participation Group, The Johns Hopkins University, 
Los Alamos National Laboratory, the Max-Planck Institute for Astronomy 
(MPIA), the Max-Planck Institute for Astrophysics (MPA), New Mexico 
State University, University of Pittsburgh, Princeton University, 
the United States Naval Observatory, and the University of Washington.
\end{acknowledgements}



\begin{thebibliography}{}

\bibitem[2003]{abazajian} Abazajian, K., et al. 2003, AJ, submitted
(astro-ph/0305492)

\bibitem[2002]{alcala} Alcal\'a, J.M., Radovich, M., Silvotti, R., Pannella, M., Capaccioli, M., \& Longo, G. 2002, Proceedings of the SPIE, 4836, 406  

\bibitem[1995]{andersen} Andersen, M.I., Freyhammer, L., \& Storm, J. 1995, in P. Benvenuti (ed.), Calibrating and understanding HST and ESO instruments, ESO Conference and Workshop
Proceeding, 53, 87

\bibitem[1998]{baade} Baade, D., et al. 1998, ESO Messenger, 93, 13 

\bibitem[1999]{baade99} Baade, D., et al. 1999, ESO Messenger, 95, 15

\bibitem[2001]{clowe} Clowe, D., \& Schneider, P. 2001, A\&A 379, 384

\bibitem[1978]{cousins} Cousins, A.W.J. 1978, MNASSA, 37, 8

\bibitem[1995]{fuk96} Fukugita, M., Ichikawa, T., Gunn, J.E., Doi, M., Shimasaku, K., \& Schneider, D.P. 1996, AJ, 111, 1748 

\bibitem[2002]{girardi} Girardi, L., Bertelli, 
G., Bressan, A., Chiosi, C., Groenewegen, M.~A.~T., Marigo, P., Salasnich, 
B., \& Weiss, A.\ 2002, A\&A, 391, 195

\bibitem[2001]{grebel2001} Grebel, E.~K.\ 2001, Reviews of 
Modern Astronomy, 14, 223 

\bibitem[2001]{gunn} Gunn, J.E., et al. 1998, AJ, 116, 3040

\bibitem[2001]{hogg} Hogg, D.W., Schlegel, D.J., Finkbeiner, D.P., \& Gunn, J.E. 2001, AJ, 122, 2129

\bibitem[1953]{johnson} Johnson, H.L., \& Morgan, W.W. 1953, ApJ, 117, 313

\bibitem[2003]{koch} Koch, A. 2003, Diploma Thesis, University of Heidelberg

\bibitem[1992]{landolt} Landolt, A.U. 1992, AJ, 104, 340

\bibitem[2001]{lupton} Lupton, R., Gunn, J. E., Ivezi\'c, Z., Knapp, G. R., 
Kent, S., \& Yasuda, N. 2001, in ASP Conf. Ser. 238, Astronomical Data 
Analysis Software and Systems X, ed. F.R. Harnden, Jr., F.A. Primini, 
and H.E. Payne (San Francisco: ASP), 269

\bibitem[2001]{manfroid_a} Manfroid, J., Selman, F., \& Jones, H. 2001a, ESO Messenger, 104, 16 

\bibitem[2001]{manfroid_b} Manfroid, J., Royer, P., Rauw., \& Gosset, E. 2001b, in Astronomical Data Analysis Software and Systems X, ASP Conf.\ Ser.\ Vol.\ 238,
ed. F.R. Harnden, Jr., F.A. Primini, and H.E. Payne (San Francisco: ASP), 373   

\bibitem[2001]{odenkirchen} Odenkirchen, M., et al. 2001, ApJ 548, L165

\bibitem[2003]{odenk03} Odenkirchen, M., et al. 2003, AJ, submitted (astro-ph/0307446)

\bibitem[1995]{roberts} Roberts, W.J., \& Grebel, E.K. 1995, A\&AS, 109, 313

\bibitem[1993]{schechter} Schechter, P.L., Mateo, M., \& Saha, A. 1993, PASP, 105, 1342

\bibitem[2002]{smith} Smith, J.A., et al. 2002, AJ, 123, 2121 

\bibitem[2000]{Stetson} Stetson, P.B., 2000, PASP, 112, 925

\bibitem[2002]{stoughton} Stoughton, C., et al. 2002, AJ, 123, 485

\bibitem[2001]{szalay} Szalay, A., et al. 2001, MSR-TR-2001-104

\bibitem[2000]{york} York, D.G., et al. 2000, AJ, 120, 1579
 
\end{thebibliography}
\end{document}